\documentclass[twocolumn,amsmath,amssymb,prb]{revtex4}
\usepackage{graphicx}% Include figure files
\usepackage{bm}% bold math
\begin{document}

\title{Two-dimensional magnetoexcitons in the presence of spin-orbit coupling} 
\author{Oleg Olendski, Quinton L. Williams, and Tigran V. Shahbazyan}
\affiliation{Department of Physics, Jackson State University, P.O. Box
  17660, Jackson, MS 39217 USA} 

\begin{abstract}
We study theoretically the effect of spin-orbit coupling on quantum well excitons in a strong magnetic field. We show that, in the presence of an \emph{in-plane} field component, the excitonic absorption spectrum develops a double-peak structure due to hybridization of bright and dark magnetoexcitons. If the Rashba and Dresselhaus spin-orbit constants are comparable, the magnitude of splitting can be tuned in a wide interval by varying the azimuthal angle of the in-plane field. We also show that the interplay between spin-orbit and Coulomb interactions leads to an anisotropy of exciton energy dispersion in the momentum plane. The results suggest a way for direct optical measurements of spin-orbit parameters.
\end{abstract}
%\pacs{PACS numbers: 71.10.Ca, 71.45.-d, 78.20.Bh, 78.47.+p}
\maketitle

%%%%%%%%%%%%%%%%%%%%%%%%%%%%%%%%%%
\section{introduction}
\label{intro}

The role of spin-orbit (SO) interactions in magnetooptics has been studied starting with the original work of Rashba.\cite{rashba-ftt60} Most of the theoretical work was devoted to the effect of the SO-induced nonparabollicity on cyclotron resonance. In quantum wells (QWs), the anticrossings of Landau levels (LLs) due to SO coupling lead to an intricate structure of the cyclotron resonance  lineshape due to the interplay of Coulomb and SO interactions in two-dimensional (2D) electron gas. \cite{falko-prb92,falko-prl93,lipparini-prb04}  In a strong \emph{tilted} magnetic field, such anticrossings occur when an in-plane component of magnetic field is tuned to bring the Zeeman-split adjacent LLs into resonance. While measurements of SO-induced beats of Shubnikov-de-Haas oscillations have long become a standard method for determining SO constants in QWs,\cite{winkler-book} there are relatively few direct observations of SO effects in optical spectroscopy; those include asymmetric spin-flip Raman scattering \cite{richards-prb99} and splitting of the cylotron resonance absorption peak.\cite{manger-prb01,vasilev-jetpl04}

In this paper, we study the effect of SO coupling on QW excitons in a strong \emph{tilted} magnetic field. 2D magnetoexcitons (MXs) are ideal objects for studying the Coulomb interaction effects.\cite{lozovik-jetp80,dzyubenko-ftt84,paquet-prb85} For sufficiently high fields, when the characteristic Coulomb energy, 
\begin{equation}
\label{E0}
E_{0}=\sqrt{\frac{\pi}{2}}\frac{e^{2}}{\kappa l}, 
\end{equation}
is smaller than the single-particle cyclotron energy, $\omega_{c}$, the relative degrees of motion are essentially frozen and Coulomb interactions play the dominant role (here $l$ is the magnetic length corresponding to the normal field component and $\kappa$ is the dielectric constant). For example, while for small center-of-mass (CM) momenta ${\bf p}$ the MX dispersion is quadratic, the MX mass is much heavier than that of the consituent electron and hole by the factor $\omega_{c}/E_{0}\gg 1$, as measured in coupled QW experiments.\cite{butov-prl01} The dominant role of Coulomb correlations is also apparent in non-Markovian ultrafast dynamics of MXs in the non-linear optical response.\cite{mukamel-prb98,shahbazyan-prb00}

\begin{figure}[t]
\vspace{10mm}
\centering
\includegraphics[width=0.5\columnwidth]{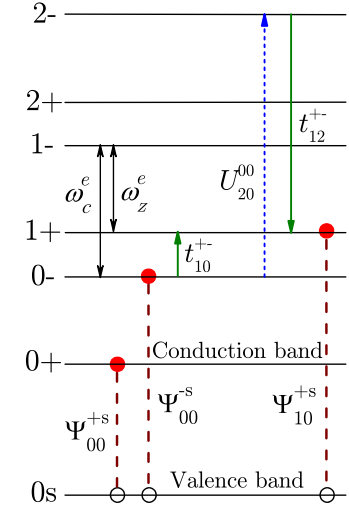}
\caption{\label{fig:LL} 
(Color online)
Schematic representation of MX states, $\Psi_{00}^{\pm s}$ (to the left), excited between $n=0$ LLs with right/left polarized light ($s=\pm$) in a tilted field. Near the resonance, Eq.~(\ref{MX-res}), states $\Psi_{00}^{- s}$ and $\Psi_{10}^{+ s}$ (to the right) are hybridized via electron SO coupling (in first order) and SO-Coulomb coupling (in second order).}
\end{figure}

In the absence of SO coupling, optically active, or bright, MXs (with spin projection $\pm1$) are those with constituent electron and hole at the $n$th level of their respective Landau ladders.\cite{bauer-prb88} The SO interaction mixes bright and dark MXs with different orbital and spin content through the SO coupling of single-particle LLs (see Fig. \ref{fig:LL}). Such a mixing is strong if the corresponding MX energies are brought close to each other, e.g., with increasing \emph{tilt angle}, $\theta$. This exciton resonance condition differs from that for electron spin resonance by the difference between MXs Coulomb binding energies. Importantly, such exciton  resonance can also occur at \emph{finite} CM momenta ${\bf p}$; this drastically changes the MX dispersion, as discussed below.

There are two distinct types of SO couplings, one originating from bulk inversion asymmetry (Dresselhaus coupling) and the other one from structural inversion asymmetry along the growth direction (Rashba coupling), that cause the admixture of orbital states with opposite spins. An important distinction between electronic Rashba and Dresselhaus SO terms is their different symmetry properties. The former possesses an in-plane rotational symmetry, while the latter does not.\cite{winkler-book} This lack of rotational invariance leads to an {\em in-plane momentum azimuthal anisotropy} in the presence of {\em both} SO terms\cite{eppenga-prb88,silva-prb92,averkiev-prb99} that was recently reported in transport\cite{marcus-prl03,ganichev-prl04} and spin relaxation\cite{winkler-prb03,averkiev-prb06} experiments in QWs. In magnetic field, where single-electron energy spectrum is dispersionless, the {\em interference} between Rashba and Dresselhaus terms leads to a dependence of SO matrix elements  on the \emph{in-plane magnetic field orientation}, $\varphi$.\cite{falko-prb92} In quantum dots, such a dependence results in a modulation of spin relaxation rate for different orientations of the in-plane field. \cite{golovach-prl04,falko-prl05,konemann-prl05,stano-prl06,olendski-prb07}

We find that the effect of SO coupling on MXs is twofold. First, the SO mixing of MX states causes anticrossings of MX energy levels with changing $\theta$. For zero CM momentum, ${\bf p}=0$, such anticrossings lead to a splitting of excitonic absorption peak, with peak-to-peak separation given by \emph{single-particle} SO anticrossing gap, $\Delta_{0}$. The splitting can be changed in a wide range by varying $\varphi$. This allows to distinguish the SO-induced MX anticrossings from those due to other mechanisms such as, e.g., heavy-light hole mixing in valence band or orbital effect of strong in-plane field.\cite{yang_sham-prb87,bauer-prb88,jho-prb05} Importantly, the angular dependence of the absorption peak lineshape would provide an independent way for direct measurements of both Rashba and Dresselhaus SO parameters in optical spectroscopy experiments. Our numerical calculations show that the splitting should be easily observed in exciton absorption experiments in narrow-gap semiconductor QWs, such as InSb, that are characterized by relatively large electron SO coupling.\cite{santos-prb01,khodaparast-prb04}

The second effect of SO coupling is to alter the MX dispersion. At fixed values of $\theta$, the MX dispersions experience anticrossing with changing \emph{CM momentum}, ${\bf p}$. As a result, at finite momentum, the dispersion curves are separated by the anticrossing gap, $\Delta_{\bf p}$. Remarkably, the interplay between SO and Coulomb interactions leads to the \emph{MX momentum anisotropy} of $\Delta_{\bf p}$ and hence of the MX dispersion. Furthermore, the MX energy landscape in the ${\bf p}$-plane depends on the in-plane field orientation, $\varphi$. In particular, for $\varphi\pm \pi/4$, the locations of extrema in  $\Delta_{\bf p}$ and, accordingly, the pattern of constant energy lines are insensitive to the values of SO parameters, while for all other $\varphi$ the pattern of equipotentials is SO-specific.

The paper is organized as follows. In Sec.~\ref{sec:so} we describe electron and hole states in a tilted field in the presence of SO coupling, while the corresponding exciton states are described in Sec.~\ref{sec:MX}. The exciton absorption and energy dispersion, together with numerical calculations, are discussed in Secs.~\ref{sec:absorption} and \ref{sec:dispersion}, respectively. Section \ref{sec:conc} concludes the paper.

\section{2D Electronic states in a tilted magnetic field in the presence of spin-orbit coupling}
\label{sec:so}

We start with the electronic spectrum in a QW in the presence of SO interactions subjected to a tilted magnetic field, ${\bf B}={\bf B}_\perp+{\bf B}_{||}=B\bigl(\hat{\bf x}\sin\theta\cos\varphi+\hat{\bf y}\sin\theta\sin\varphi+\hat{\bf z}\cos\theta\bigr)$, where $\theta$ is the tilt angle and  $\varphi$ is the asimuthal angle with respect to crystallographic axes of the [001] plane. We consider the QW to be sufficiently narrow and the effect of an in-plane field component on orbital motion to be negligibly small.  The electron Hamiltonian in the conduction band,  $H_{e}=H_{0}^{e}+H_Z^{e}+H_{so}^{e}$, is comprised of orbital term, $H_{0}^{e}={\bm \pi}^{2}/2m_{e}$, Zeeman term, $H_Z^{e}=\frac{1}{2}g_{e}^\ast\mu_B {\bm \sigma}\cdot {\bf B}$, and SO term, $H_{so}^{e}=H_{R}^{e}+H_{D}^{e}$. Here  $m_{e}$ and $g_{e}^\ast$ are the electron effective mass and $g$-factor, respectively, $\mu_B$ is the Bohr magneton, ${\bm \sigma}$ is the Pauli matrices vector, and  ${\bm \pi}=-i\nabla + e{\bf A}$ is the in-plane momentum [we use the Landau gauge, ${\bf A}=(0, xB_\perp)$, and set $\hbar=1$ throughout]. Two contributions to $H_{so}^{e}$ are Rashba and Dresselhaus terms, $H_{R}^{e}=i\alpha(\sigma_{+}\pi_{-}-\sigma_{-} \pi_{+})$ and $H_{D}^{e}=\beta(\sigma_{+}\pi_{+}+\sigma_{-} \pi_{-})$, respectively, $\alpha$ and $\beta$ being the corresponding SO constants, where $\pi_{\pm}=\pi_{x}\pm i\pi_{y}$ and $\sigma_{\pm}=(\sigma_{x}\pm i\sigma_{y})/2$. Hole states in the valence band have, in general, a more complicated structure due to the mixing of heavy hole (HH) and light hole (LH) states by an in-plane magnetic field.\cite{bauer-prb88}  However, for \emph{narrow} QW and \emph{low} LLs that we are interested in, the HH states are well separated from the LH band so this mixing is weak.\cite{winkler-book} In this case, the in-plane HH $g$-factor is negligible, i.e., the total momentum $\hat{\bf J}$ is quantized along the $z$-axis, $J_z=\pm 3/2$, even in the presence of an in-plane field component ${\bf B}_{||}$. The HH Zeeman Hamiltonian, $H_Z^{h}$, therefore has the form  $H_Z^{h}=-\frac{1}{2}g_h^\ast\mu_B B_\perp\sigma_z$, where the eigenvalues of $\sigma_z$ correspond to the two projections $J_z$ and $g_h^\ast$ is the effective HH $g$-factor in the growth direction. Accordingly, we adopt a simple one-band HH Hamiltonian, $H_h=H_{0}^{h}+H_Z^{h}+H_{so}^{h}$, where   $H_{so}^{h}$ is the SO term that is cubic in momentum, $H_{so}^{h}= i\tilde{\alpha}(\sigma_{+}\pi_{-}^{3}-\sigma_{-} \pi_{+}^{3})- \tilde{\beta}(\sigma_{+}\pi_{-}\pi_{+}\pi_{-}+\sigma_{-} \pi_{+}\pi_{-}\pi_{+})$, $\tilde{\alpha}$ and $\tilde{\beta}$ being valence band SO couplings.\cite{bulaev-prl05} 

In contrast, in the conduction band, the ``natural'' spin quantization axis is along the total field ${\bf B}$. At the same time, the above form of electronic SO terms applies when \emph{x}, \emph{y}, and \emph{z}-directions are aligned with the sample crystallographic axes. Therefore, correct expressions for the SO terms in a \emph{tilted} field are obtained upon rotation of spin operators to align the spin-quantization axis with the total field:\cite{olendski-prb07} $\sigma_{\pm}\rightarrow e^{\pm i\varphi}\left[ \sigma_{\pm}\cos^{2}(\theta/2) - \sigma_{\mp}\sin^{2}(\theta/2) +(\sigma_{z}/2)\sin \theta \right] $, and $\sigma_z\rightarrow \sigma_z\cos\theta -(\sigma_{+} + \sigma_{-})\sin\theta$. In this  basis, $H_{so}^{e}$ reads
\begin{eqnarray}
\label{so-new}
H_{so}^{e}=\frac{\pi_{+}}{2}\Bigl[ \sigma_{+}(\gamma_{+}+\gamma_{-}\cos\theta)-\sigma_{-}(\gamma_{+}-\gamma_{-}\cos\theta)
\nonumber\\
+ \sigma_{z}\gamma_{-}\sin\theta \Bigr] + {\rm h.c.} ,
\end{eqnarray}
where $\gamma_{\pm}(\varphi)=\beta e^{i\varphi}\pm i\alpha e^{-i\varphi}$. In a tilted field, no analytical expression exists for eigenstates of $H_{e}$, but we only need  matrix elements $t_{nn'}^{ss'}\equiv \langle ns|H_{so}^{e}|n's'\rangle$ between the eigenstates of $H_{0}^{e}+H_Z^{e}$. The latter are given by products of Landau wave-functions and two-component spinors, $\psi_{p_yn}^{s}({\bf r})= \psi_{p_yn}({\bf r})\chi^s_0$ with $\tilde{\chi}_0^+=(1\,0)$ and $\tilde{\chi}_0^-=(0\,1)$; the corresponding energies are
\begin{eqnarray}
\label{e-energy}
 E_{ns}^{e}=\omega_{c}^{e}(n+1/2)- s \omega_{z}^{e}/2,
\end{eqnarray}
where $n=0,1,_{\cdots}$, $s=\pm 1$ is LL number and $\omega_{c}^{e}=eB_\perp/m_e$ and  $\omega_{z}^{e}=-g_e^\ast \mu_B B$ are cyclotron and Zeeman frequencies, respectively (hereafter, we asume negative $g$-factor). For adjacent LLs,  using $\langle n+1|\pi_+|n\rangle=i\sqrt{2(n+1)}/l$, Eq.~(\ref{so-new}) yields
\begin{eqnarray}
\label{matrix-all}
t_{n+1, n}^{\pm\mp}= \pm \frac{i}{l}\sqrt{\frac{n+1}{2}}\, \Bigl[\gamma_{+}(\varphi)\pm \gamma_{-}(\varphi)\cos\theta\Bigl],
\nonumber\\
t_{n, n+1}^{\pm\mp}= \pm \frac{i}{l}\sqrt{\frac{n+1}{2}}\, \Bigl[\gamma_{+}^{\ast}(\varphi)\mp \gamma_{-}^{\ast}(\varphi)\cos\theta\Bigl].
\end{eqnarray}
In a strong field, the characteristic SO energy is small compared to the level separation, $|\gamma_{\pm}|/l \ll\omega_{c}^{e}$, and, accordingly, the SO-induced level admixture is, in general, weak. However, the mixing gets strongly enhanced when the spacing between adjacent LLs with opposite spins is reduced, e.g., by varying the Zeeman energy with the tilt angle $\theta$ (see Fig.~\ref{fig:LL}). In this case, the SO coupling leads to level anticrossing at $|\omega_{c}^{e}-\omega_{z}^{e}| \sim |\gamma_{\pm}|/l$. The anticrossing gap between, e.g., lowest resonant levels, $\Delta_{0}=2|t_{10}^{+-}|=2|\langle 1+|H_{so}^{e}|0-\rangle|$, 
\begin{eqnarray}
\label{gap}
\Delta_{0}
%|\langle 0-|H_{so}|1+\rangle|
=2\sqrt{\frac{2\alpha^2}{l^2}\sin^4\frac{\theta}{2}+
\frac{2\beta^2}{l^2}\cos^4\frac{\theta}{2}
%\qquad \qquad \qquad 
%\nonumber\\
+ \frac{\alpha\beta}{l^2}\sin^2\theta\sin 2\varphi},
\end{eqnarray}
depends on the orientation of the in-plane field component,\cite{falko-prb92} i.e., with $\theta$ fixed by the resonance condition, $\omega_{c}^{e}=\omega_{z}^{e}$,  the gap  varies with azimuthal angle $\varphi$. Note that for $\varphi=- \pi/4$ and $\beta/\alpha=\tan^2(\theta/2)$, there is \emph{destructive interference} between the two SO terms in the matrix elements $t_{n+1, n}^{\pm\mp}$, and $\Delta_{0}$ {\em vanishes} in the first order in SO coupling; higher-order corrections involving SO coupling to upper LLs are suppressed as $(|\gamma_{\pm}|/l\omega_{c}^{e})^2\ll 1$. For $t_{n, n+1}^{\pm\mp}$, the above condition applies upon replacement  $\alpha\leftrightarrow\beta$.

In contrast, in the valence band, the negligible value of in-plane HH $g$-factor precludes occurence of similar resonances between neighboring LL's.  For $g_{h}^{\ast}<0$ (e.g., in InAs or InSb), the lowest state  $| 0-\rangle$ is only weakly coupled, via $H_{so}^{h}$, to $| 1+\rangle$ (via Dresselhaus term) and $| 3+\rangle$ (via Rashba term) states, while the upper state $| 0+\rangle$  is not coupled to other LLs; for  $g_{h}^{\ast}>0$  (e.g., in GaAs),  $| 0+\rangle$ is the lowest state. 

%%%%%%%%%%%%%%%%%%%%%%%%%%%%%%%%%%%%%%%%% 
\section{Magnetoexciton states in the presence of spin-orbit coupling}
\label{sec:MX}

We now turn to exciton states in a tilted field described by the Hamiltonian $H=H_{e}+H_{h}+H_{eh}$, where $H_{eh}$ is the Coulomb interaction. We assume that the perpendicular field component is sufficiently strong, $\omega_{c}^{e} \gg E_{0}\gg |\gamma_{\pm}|/l$, so that Coulomb-induced inter-LL transitions are relatively weak. In the absence of SO coupling, exciton states are expressed via free electron-hole (\emph{e-h}) basis functions, $\Psi^{ss^\prime}_{{\bf p}nm}({\bf r},{\bf r}')=\Psi_{{\bf p}nm}({\bf r},{\bf r}')S^{ss'}$, where ${\bf r}$ and ${\bf r}'$ are electron and hole coordinates, respectively. The orbital part, corresponding to an electron at the $n$th and a hole at the $m$th LLs, is given by \cite{lozovik-jetp80,kallin-prb84,paquet-prb85}
\begin{equation}
\label{Psi-exc}
\Psi_{{\bf p}nm}({\bf r}, {\bf r}')
=\frac{1}{L}
e^{i{\bf p\cdot R}-iXy/l^2}
\varphi_{nm}(\tilde{\bf r}+l^2{\bf p} \times \hat{\bf z}),
\end{equation}
where ${\bf p}$ is the CM momentum of an \emph{e-h} pair, $\tilde{\bf r}={\bf r}-{\bf r}'$,  ${\bf R}=({\bf r}+{\bf r}')/2$ are the relative and average coordinates, respectively ($L$ is system size), and 
\begin{equation}
\label{phi-exc}
\varphi_{nm}(z)=\sqrt{\frac{m!}{n!}}\Bigl(\frac{iz}{\sqrt{2}l}\Bigr)^{n-m}L_{m}^{n-m}\Bigl(\frac{|z|^2}{2l^2}\Bigr)\frac{e^{-|z|^2/4l^2}}{\sqrt{2\pi l^2}} 
\end{equation}
is the relative motion wave-function [$L_n^\alpha\left(x\right)$ is the Laguerre polynomial,  $z=x+iy$]. The spin part is a diadic product  of electron and HH spinors, $S^{ss^\prime}=\chi_{0e}^s \otimes \chi_{0h}^{s^\prime}$, with electron and hole spin-quantization axes along ${\bf B}$ and ${\bf B}_{\perp}$, respectively. For $E_{0}/\omega_{c}^{e}\ll 1$,  the MX eigenstates are obtained perturbatively in the basis of Eq.~(\ref{Psi-exc}). In the first order, i.e. neglecting inter-LL transitions, the wave-function does not change, while the MX energy is given by 
\begin{equation}
\label{MX-energy}
E_{nm}^{ss'}(p)=E_{g}+E_{ns}^{e}+E_{ms'}^{h} +  U_{nn}^{mm}(p),
\end{equation}
where $E_{ns}^{e,h}$ are given by Eq.~(\ref{e-energy}), $E_g$ is the bandgap, and $U_{nn}^{mm}(p)$ is the diagonal matrix element of Coulomb potential $V({\bf r}-{\bf r}')=e^{2}/\kappa|{\bf r}-{\bf r}'|$,
\begin{equation}
\label{coulomb}
U_{nn'}^{m m'}\left(\bf p\right)=-\int d{\bf r}d{\bf r}'\Psi^{\ast}_{{\bf p}nm}({\bf r},{\bf r}') V({\bf r}-{\bf r}') \Psi_{{\bf p}n'm'}({\bf r},{\bf r}'), 
\end{equation}
with lower and upper indices refering, respectively, to electron and hole quantum numbers.

The  SO coupling causes the admixture of MX states with different orbital and spin content. The corresponding matrix element, 
\begin{equation}
\label{SO-matrix}
T_{\nu\nu'}^{\lambda\lambda'}({\bf p}) \equiv \langle {\bf p}\nu\lambda|\bigl(H_{so}^{e}+H_{so}^{h}\bigr)|{\bf p}\nu'\lambda'\rangle, 
\end{equation}
with $\nu$ and $\lambda=(ss')$ denoting sets of orbital and spin indices, respectively, is a sum of electron and hole SO contributions. In the first order in $E_{0}/\omega_{c}^{e}$, as Coulomb-induced inter-LL transitions are suppressed, the orbital part of $|{\bf p}\nu\lambda\rangle$ coincides with Eq.~(\ref{Psi-exc}). In this case, the excitonic SO transition operator reduces to the sum of tensor products $\hat{T}=\hat{t^{e}}\otimes \hat{I^{h}} + \hat{t^{h}}\otimes\hat{I^{e}}$, where $\hat{t}^{e,h}$ are single-particle SO transition operators, and $\hat{I}^{e,h}$ are unit tensors in corresponding orbital and spin indices. 

There are four MX states at the lowest LL corresponding to all possible orientations of electron spin and hole total momentum, that are mixed with higher energy states  by SO coupling  in conduction and valence band. Note that, due to large energy separation between the corresponding states, the smallness of in-plane HH $g$-factor results in only weak SO-mixing in the presence of in-plane field. Therefore, in the following we consider only the effect of conduction band SO-mixing. The state $\Psi^{+s}_{{\bf p}00}$, with $s=\pm$ for either hole polarization, is weakly coupled, via electronic SO matrix elements Eq.~(\ref{matrix-all}), to the state $\Psi^{-s}_{{\bf p}10}$ that lies significantly higher in energy due to a large Zeeman splitting $\omega_{z}^{e}$ (see Fig.~\ref{fig:LL}). At the same time, the state $\Psi^{-s}_{{\bf p}00}$ couples to  $\Psi^{+s}_{{\bf p}10}$; their energy separation,  $\delta_{p}$, is given by
\begin{equation}
\label{delta}
\delta_p=E_{10}^{+s}(p)-E_{00}^{-s}(p)=\omega_{c}^{e}-\omega_{z}^{e} +U_{11}^{00}(p)-U_{00}^{00}(p),
\end{equation}
where  
\begin{align}
\label{U00}
U_{00}^{00}(p)=&-E_{0}e^{-x}I_0(x),
\nonumber\\
U_{11}^{00}(p)=&-E_{0}e^{-x}\biggl[\left(\frac{1}{2}+x\right )I_0\left(x\right) -xI_1(x)\biggr],
\end{align}
%
%%
%\begin{align}
%\label{U00}
%U_{00}^{00}(p)=&-E_{0}e^{-p^2l^2/4}I_0\left(\frac{p^2l^2}{4}\right),
%\nonumber\\
%U_{11}^{00}(p)=&-E_{0}e^{-p^2l^2/4}\biggl[\left(\frac{1}{2}+\frac{p^2l^2}{4}\right )I_0\left(\frac{p^2l^2}{4}\right) 
%\nonumber\\
%& \qquad \qquad \qquad \qquad ~
%-\frac{p^2l^2}{4}I_1\left(\frac{p^2l^2}{4}\right)\biggr],
%\end{align}
%%
are the relevant Coulomb matrix elements with $x=p^2l^2/4$ [$I_n(x)$ is the modified Bessel function].  Note that, for MXs, the resonance condition $\delta_p=0$ is Coulomb-shifted from single-particle one; in particular, for ${\bf p}=0$, it reads 
\begin{equation}
\label{MX-res}
\omega_{z}^{e}-\omega_{c}^{e}=E_0/2.
\end{equation}
In the in-plane field domain where $|\delta_{p}|\sim |t_{10}^{+-}|$, the admixture is strong and the new eigenenergies are 
\begin{equation}
\label{eigen-energy}
E^{s}_{\pm}(p)=\frac{1}{2}
\Bigl[E^{-s}_{00}\left(p\right)+E^{+s}_{10}\left(p\right)\pm\sqrt{\delta_p^2+\Delta^2}\,\Bigr],
\end{equation}
where, in the absence of inter-LL transitions, $\Delta=\Delta_{0}$ coincides with single-particle anticrossing gap, Eq.~(\ref{gap}), and is $p$-independent. The corresponding eigenstates are superpositions of unperturbed exciton states with close energies,
\begin{equation}
\label{eigen-function}
\Psi_{{\bf p}\pm}^{s}\left({\bf r},{\bf r}'\right)
=a_{p}^{\pm}\Psi_{{\bf p}00}^{-s}\left({\bf r},{\bf r}'\right)
+b_{p}^{\pm}\Psi_{{\bf p}10}^{+s}\left({\bf r},{\bf r}'\right),
\end{equation}
where the coefficients $a_{p}^{\pm}$ and $b_{p}^{\pm}$ are determined by diagonalizing the full Hamiltonian $H=H_e+H_h+H_{eh}$,
\begin{eqnarray}
\label{coeff}
a_p^{-}=b_p^{+}=\frac{1}{\sqrt{1+e^{-2\beta_p}}},
\,
a_p^{+}=-b_p^{-\ast}=\frac{e^{i\eta}}{\sqrt{1+e^{2\beta_p}}}.
\end{eqnarray}
Here $\eta=\arg(t_{10}^{+-})$ is the phase of the electron SO matrix element, and the 
parameter $\beta_p$, defined by $\sinh\beta_p=\delta_p /\Delta$, is the detuning in units of the anticrossing gap that characterizes the proximity to the resonance. Note that outside of the resonance region, $|\delta_p|\gg \Delta$ (but still $|\delta_p|\ll \omega_{c}^{e}$), we have $a_p$ and $b_p$ equal 0 or 1 so that the two excitons are almost decoupled.

%%%%%%%%%%%%%%%%%%%%%%%%%%%%%%%%%%%%

\section{Magnetoexciton absorption}
\label{sec:absorption}

A circularly-polarized light incident normal to the plane can excite only \emph{e-h} pairs with total spin projection $\sigma=\pm 1$ for right/left polarized photon, respectively. The optically active excitations with $\sigma=\pm 1$ are an electron and a hole at the  $n$th LLs with  $J_{z}=\pm 3/2$ for hole and $s_{z}=\mp 1/2$ for electron.\cite{bauer-prb88} The corresponding wavefunctions are $\Psi_{n}^{\sigma}({\bf r},{\bf r}')=\Psi_{{\bf 0}nn}({\bf r},{\bf r}')S^{\sigma}$, where the orbital part is taken at ${\bf p}=0$ due to negligible momentum of incident photon, and in the spin part, $S^{\pm}=\chi_{0h}^{\pm}\otimes \chi_{e}^{\mp}$,  the spinor $\chi_{e}$ stands for electron spin projection perpendicular to the plane. In the basis with electron spin-quantization axis along total field ${\bf B}$, we have $\chi_{e}^{+}=\chi_{0e}^{+}\cos(\theta/2) - \chi_{0e}^{-}\sin(\theta/2)$ and $\chi_{e}^{-}=\chi_{0e}^{+}\sin(\theta/2)+\chi_{0e}^{-}\cos(\theta/2)$.  In the following, we restrict ourselves to optical excitations with energies close to $n=0$ LLs. The excitonic absorption coefficient has the form 
\begin{equation}
\label{absorp}
A^{\sigma}(\omega)\propto \sum_{\alpha}|C_{\alpha}^{\sigma}|^2 \delta(\omega-E_{\alpha}),
\end{equation}
where the sum runs over MX eigenstates with energies $E_{\alpha}$ and  ${\bf p}=0$ ($\alpha$ incorporates both orbital and spin indices); the corresponding oscillator strengths are given by 
\begin{equation}
\label{oscill}
C_{\alpha}^{\sigma}=\mu \int d{\bf r} d{\bf r}'\Psi_{\alpha}^{\dagger}({\bf r},{\bf r}') \Psi_{0}^{\sigma}({\bf r},{\bf r}'), 
\end{equation}
$\mu$ being the interband dipole matrix element.
 
Consider first absorption of right circularly-polarized light. The state $\Psi_{0}^{+}=\Psi^{-+}_{{\bf 0}00}\cos(\theta/2)+\Psi^{++}_{{\bf 0}00}\sin(\theta/2)$ is not an eigenstate of the system because of SO-mixing of constituent exciton states with upper LLs. The state $\Psi^{++}_{{\bf 0}00}$ is only weakly coupled to $\Psi^{-+}_{{\bf 0}10}$, as mentioned above, and is, in a good approximation, an eigenstate contributing oscillator strength $\mu^{2}\sin^{2}(\theta/2)$ into the sum (\ref{absorp}). Correspondingly, the absorption spectrum exhibits a peak at frequency $ E^{++}_{00}(0)$ that appears only in a tilted field. On the other hand, in the resonance region, the state $\Psi^{-+}_{{\bf 0}00}$ is strongly coupled to $\Psi^{++}_{{\bf 0}10}$ (see Fig.~\ref{fig:LL}), so that eigenstates are $\Psi_{{\bf 0}\pm}^{+}$, given by Eq.~(\ref{eigen-function}), yielding
\begin{equation}
\label{oscill-right}
|C_{\pm}^{+}|^2=
\mu^2 \cos^2(\theta/2)|a_{0}^{\pm}|^2=
\frac{\mu^2\cos^2(\theta/2)}{1+e^{\pm 2\beta_0}},
\end{equation}
with $\sinh\beta_0=\delta_0/\Delta_{0}=(\omega_{z}^{e}-\omega_{c}^{e}-E_0/2)/\Delta_{0}$.
As result, the absorption spectrum exhibits \emph{double-peak} structure at energies $E^{+}_{\pm}(0)$ given by Eq.~(\ref{eigen-energy}).

Similarly, for left-polarized absorption, the bright state is decomposed as $\Psi_{00}^{-}=\Psi^{+-}_{{\bf 0}00}\cos(\theta/2)-\Psi^{--}_{{\bf 0}00}\sin(\theta/2)$. The state $\Psi^{+-}_{{\bf 0}00}$ is weakly coupled to higher-energy states, and so contributes oscillator strength $\mu^{2}\cos^{2}(\theta/2)$ into the sum (\ref{absorp}) corresponding to the absorption peak at frequency  $E^{+-}_{{\bf 0}00}$. At the same time, in the resonance region, the state $\Psi^{--}_{{\bf 0}00}$ is strongly mixed with  $\Psi^{+-}_{{\bf 0}10}$, and we obtain
\begin{equation}
\label{oscill-left}
|C_{\pm}^{-}|^2=
\mu^2 \sin^2(\theta/2)|a_{0}^{\pm}|^2=
\frac{\mu^2\sin^2(\theta/2)}{1+e^{\pm 2\beta_0}}.
\end{equation}
The corresponding absorption spectrum lineshape develops a double-peak  structure at  energies $E^{-}_{\pm}(0)$. The peak amplitude differs by the factor $\tan^2\left(\theta/2\right)$ from that of its right-polarized counterpart; the absorption is non-zero only in the presence of in-plane field component.

Thus, in a tilted field, the SO coupling leads to a splitting of the MX absorption peak when the energies of  MXs with different spin content are brought into resonance, e.g., by varying in-plane field component, Eq.~(\ref{MX-res}). The peak-to-peak separation is given by the SO-induced anticrossing gap in the conduction band, Eq.~(\ref{gap}), that can be changed in a wide range by varying the in-plane azimuthal angle $\varphi$ with respect to [100] axis. The maximal and minimal values are achieved for $\varphi=\pm \pi/4$,
\begin{equation}
\label{gap-field}
\Delta_{0}^{\pm}=\frac{\sqrt{2}}{l}\left|\alpha\left(1-\frac{B_\perp}{B}\right)
\pm\beta\left(1+\frac{B_\perp}{B}\right)\right|.
\end{equation}
Remarkably, from measured values of $\Delta_{0}^{\pm}$ one can determine both the magnitudes and the relative sign of SO couplings $\alpha$ and $\beta$ [for opposite relative sign, the values (\ref{gap-field}) are achieved for $\varphi=\mp \pi/4$]. The splitting disappears, $\Delta_{0}^{-}=0$, at 
\begin{equation}
\label{interference}
\frac{\alpha}{\beta}=\frac{B+B_{\perp}}{B-B_{\perp}}
\end{equation}
and $\varphi=-\pi/4$,  corresponding to the destructive interfence between Rashba and Dresselhaus terms. 

In QWs, the relative strength of each type of spin-orbit interaction can be tuned in a wide range. The 2D Dresselhaus coupling in a narrow QW is determined mainly by its width $d$, $\beta=\gamma \bigr(\frac{\pi}{d}\bigr)^2$, where $\gamma$ is a material dependent parameter (the effect of cubic terms is relatively small). On the other hand, the Rashba coupling parameter can be changed with applied gate voltage,\cite{winkler-book} $\alpha\simeq r^{6c6c}_{41}{\cal E}_z$, where coefficient $r^{6c6c}_{41}$ is material-dependent and ${\cal E}_z$ is the electric field perpendicular to the plane. In materials with large $r^{6c6c}_{41}$ (e.g., $r^{6c6c}_{41}=523$ e\AA$^2$ for InSb\cite{winkler-book}), the above condition for  destructive interference of Rashba and Dresselhaus terms can be easily achieved. 

Our numerical calculations were performed for $d=10$ nm wide InSb QW in a tilted field whose normal component was taken to be $B_\perp =4.0$ T, corresponding to $\omega_{c}^{e}\approx 33.0$ meV and $E_{0}\approx 8.5$ meV. Since for InSb $\gamma=160$ eV{\AA}$^3$,\cite{ulloa-prb05} we have for Dresselhaus coupling $\beta\approx 157$ meV{\AA} corresponding to the characteristic SO energy of $\beta/l\approx 1.2$ meV. Other parameters for InSb used were: $m_e=0.014m_0$ ($m_0$ is free electron mass), effective electron $g$-factor $g_e^\ast=-51$, dielectric constant $\kappa=16.5$, and MX homogeneous broadening $\Gamma =1.0$ meV.\cite{santos-prb01} In order to assess the accuracy of the resonant level model, we included SO coupling between all four lowest spin-split electronic LLs but, in the anticrossing region,  detected virtually no difference  for the set of parameters used.

%\begin{widetext}
%
\begin{figure}[tb]
\centering
\includegraphics[width=0.7\columnwidth]{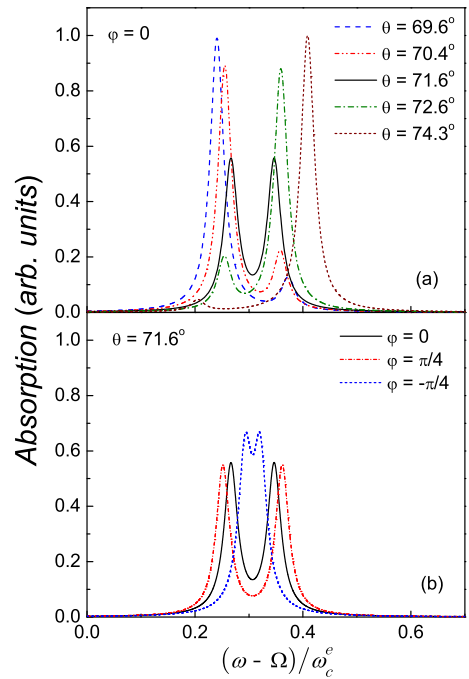}
\caption{\label{fig:absorption-1} 
(Color online)
Exciton absorption spectrum for the right circularly-polarized light for $\varphi=0$ at several values $\theta$ (a), and for $\theta=71.6^\circ$ at several values of $\varphi$ (b). Spectra are calculated with $\beta=157$ meV{\AA} and $\alpha=180$ meV{\AA}.}
\end{figure}
\begin{figure}[tb]
\centering
\includegraphics[width=0.7\columnwidth]{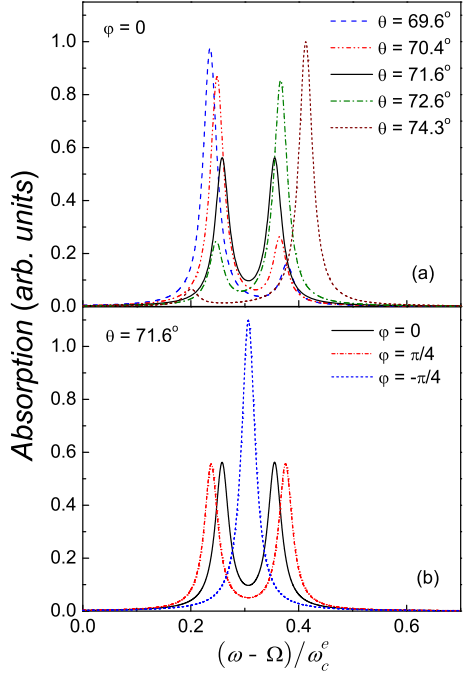}
\caption{\label{fig:absorption-2} 
(Color online) Same as Fig.~\ref{fig:absorption-1} but with $\alpha=302$ meV{\AA} satisfying destructive interference condition, Eq.~(\ref{interference}).}
\end{figure}
%
%\end{widetext}

In Figs.~\ref{fig:absorption-1} and \ref{fig:absorption-2}, we show absorption spectra for right circularly-polarized light in the frequency range corresponding to excitation of MX comprised of a $n=0$, $J_{z}=+3/2$ LL hole and an electron hybridized between $n=0$,  $s_{z}=-1/2$ and $n=1$,  $s_{z}=1/2$ LLs. The frequency is measured relative to $\Omega=E_g+(\omega_c^h+\omega_z^h+\omega_c^e)/2$ so the lower energy single peak corresponding to excitation of $\Psi_{{\bf 0}00}^{++}$ MX state is not shown. The Rashba SO parameters are taken as $\alpha=180$ meV{\AA} for Fig.~\ref{fig:absorption-1} and $\alpha=302$ meV{\AA} for Fig.~\ref{fig:absorption-2}; the latter value corresponds to the destructive interference condition (see below). The evolution of the absorption spectra with an in-plane magnetic field at azimuthal angle value $\varphi=0$ is shown in panels (a). When the tilt angle $\theta$ lies within a narrow interval $\sim 5^{\circ}$ around the resonance value, determined by Eq.~(\ref{MX-res}), the spectrum develops a double-peak structure with maxima corresponding to the excitation of hybrid states with energies $E_{\pm}^{+}(0)$. The peaks are split symmetrically at the resonance that takes place at  $\theta=71.6^{\circ}$ ($B_{\|}/B_\perp = 3.0$) when the states $\Psi_{{\bf 0}00}^{-+}$ and $\Psi_{{\bf 0}10}^{++}$ contribute equally to the final state. As one moves away from the resonance, the double-peak structure gradually transforms into a single peak with a weak shoulder. 

The absorption spectra lineshapes exhibit strong dependence on the in-plane field orientation, $\varphi$. This is illustrated in panels (b) for two different Rashba SO parameter values. For $\alpha=302$ meV{\AA}, corresponding to destructive interference between the two SO terms, Eq.~(\ref{interference}), the splitting disappears for in-plane field orientation $\varphi=-\pi/4$, while for other values of $\varphi$ it is quite pronounced [see Fig.~\ref{fig:absorption-2}(b)]. For general values of SO coupling, the splitting is visible for \emph{all} values of $\varphi$, as shown in Fig.~\ref{fig:absorption-1}(b).  The large value of peak-to-peak separation, up to $0.2\omega_{c}^{e}\approx 6.0$ meV for $\varphi=\pi/4$, is due to the strong SO coupling in InSb. Note that, at fixed $\theta$, the SO-induced splitting increases with $B_{\perp}$.

Importantly, the magnitude of excitonic absorption peak splitting is determined solely by \emph{single-particle} SO parameters encoded in  $\Delta_{0}(\varphi)$, Eq.~(\ref{gap}). This suggests a new way for direct determination of electron SO constants from optical measurements by monitoring the evolution of double-peak structure with varying in-plane field orientation, $\varphi$. For example, from maximal and minimal peak-to-peak separation, $\Delta_{0}^{\pm}$, given by Eq.~(\ref{gap-field}), the SO coefficients are deduced as  
\begin{equation}
\label{so-values}
\Bigl\{\begin{array}{r}\alpha\\ \beta\end{array}\Bigr\}=\frac{l}{2\sqrt{2}}\frac{\Delta_{0}^{+}\pm\Delta_{0}^{-}}{1\mp B_{\perp}/B}.
\end{equation}
%

%%%%%%%%%%%%%%%%%%%%%%%%%%%%%%%%%%%%%%
\section{Energy dispersion and angular anisotropy}
\label{sec:dispersion}

We now turn to the effect of SO coupling on the MX dispersion. As in the case of absorption, the role of SO coupling becomes important near the resonance, i.e., when the energy separation between MX eigenstates, Eq.~(\ref{eigen-function}),
\begin{equation}
\label{delta-energy}
E^{s}_{+}(p)-E^{s}_{-}(p)=\sqrt{\delta_p^2+\Delta^2},
\end{equation}
becomes of the order of the characteristic SO energy:  $\delta_{p}\sim \Delta$, with $\delta_{p}$ given by Eq.~(\ref{delta}). Note, however, that the latter condition can be also achieved by changing the MX momentum at a fixed tilt angle $\theta$, in contrast to the $p=0$ case in absorption where the resonance, Eq.~(\ref{MX-res}), could be reached only by changing Zeeman energy with the in-plane field component. Thus, as $\delta_{p}$ passes through the resonance, $\delta_{p}=0$, the energy dispersions of the MX states $\Psi_{{\bf p}\pm}^{s}$ experience an anticrossing \emph{as a function of momentum}. In the \emph{absence of inter-LL transitions}, the anticrossing gap $\Delta$ coincides with the single-particle gap $\Delta_{0}(\varphi)$ and is $p$-independent, as mentioned in Sec.~\ref{sec:MX}. Consequently, in this approximation, the dispersion of the MX eigenstates  depends on the in-plane field orientation, $\varphi$, but remains isotropic with respect to the MX momentum orientation, ${\bf p}$.

Situation changes drastically when Coulomb-induced inter-LL transitions are turned on. In the absence of the SO coupling, the MX energy $E_{nm}^{ss'}$ acquires a correction, 
\begin{equation}
\label{inter-psi}
\delta E_{nm}^{ss'}(p)=\sum_{n'm'}\frac{\bigl|U_{n'n}^{m'm}({\bf p})\bigr|^{2}}{E_{nm}^{ss'}(p)-E_{n'm'}^{ss'}(p)}.
\end{equation}
In the case of $E_{0}\ll\omega_{c}^{e}$, this correction  slightly changes the energy difference $\delta_{p}$ and, accordingly, merely shifts the resonance position, $\delta_{p}=0$. On the other hand, the LL mixing gives rise to a new contribution into the SO matrix element, Eq.~(\ref{SO-matrix}), originating from the \emph{interplay between SO and Coulomb couplings}. Indeed, the corresponding MX wavefunctions,  $\Psi^{ss^\prime}_{{\bf p}nm}$, acquire a correction 
\begin{equation}
\label{delta-}
\delta \Psi^{ss^\prime}_{{\bf p}nm}=\sum_{n'm'}\frac{U_{n'n}^{m'm}({\bf p})}{E_{nm}^{ss'}(p)-E_{n'm'}^{ss'}(p)} \, \Psi^{ss^\prime}_{{\bf p}n'm'}.
\end{equation}
Then, the matrix elements of $H_{so}^{e}$ between states $\Psi^{\mp s}_{{\bf p}nn}+\delta \Psi^{\mp s}_{{\bf p}nn}$ and $\Psi^{\pm s}_{{\bf p}n+1,n}+\delta \Psi^{\pm s}_{{\bf p}n+1,n}$ take the form
\begin{eqnarray}
\label{Tn}
T_{n+1, n}^{\,\,\,\pm\,\,\,\,\mp}
=t_{n+1, n}^{\,\,\,\pm\,\,\,\,\mp}
+t_{n+1,n+2}^{\,\,\,\pm\,\,\,\,\,\,\,\mp} \,
\frac{U_{n+2,n}^{nn}}{E^{\mp s}_{n+2,n}-E^{\mp s}_{nn}}
\nonumber\\
+\frac{U_{n+1,n-1}^{nn}}{E^{\pm s}_{n+1,n}-E^{\pm s}_{n-1,n}}\,
t_{n-1, n}^{\,\,\,\pm\,\,\,\,\mp},
\end{eqnarray}
where we omitted hole indices in $T$ and neglected the higher-order corrections. The first term in the r.h.s. of Eq.~(\ref{Tn}) originates from the direct SO coupling of electronic states $|n-\rangle$ and $|n+1,+\rangle$,  given by Eq.~(\ref{matrix-all}). The second term, in turn,  describes the coupling between the same levels via a two-step process: the electron is first promoted to the $|n+2,-\rangle$  state by the \emph{hole} Coulomb potential, and then makes SO-transition down to the  $|n+1,+\rangle$ state (see Fig.~\ref{fig:LL}). The last term describes a similar process involving the $(n-1)$th LL as the intermediate state. Note that for $n=0$, the last term is absent and the SO matrix element reduces to
\begin{equation}
\label{T0}
T^{+-}_{10}\left(\bf p\right)=t^{+-}_{10}+\, t^{+-}_{12} \, \frac{U_{20}^{00}\left(\bf p\right)}{E^{-s}_{20}\left(p\right)-E^{-s}_{00}\left(p\right)},
\end{equation}
where the  Coulomb matrix element is given by
\begin{equation}
\label{U2000}
U_{20}^{00}\left(\bf p\right)=e^{2i\phi_{\bf p}}\frac{E_0}{\sqrt{2}}\left( \frac{pl}{2}\right)^2f\left(p\right).
\end{equation}
Here $\phi_{\bf p}=\arg({\bf p})$ is polar angle of the 2D \emph{exciton} momentum, and $f\left(p\right)=e^{-p^2l^2/4}
\bigl[I_0\bigl(p^2l^2/4\bigr)-\bigl[1+(2/p^2l^2)\bigr]I_1\bigl(p^2l^2/4\bigr)\bigr]$ is a scalar function of the order one normalized to $f\left(0\right)=3/4$. Importantly, although the second term in Eq.~(\ref{T0}) is parametrically small by the factor $E_{0}/\omega_{c}^{e}$, as compared to the first one, it introduces an explicit dependence on the orientation of ${\bf p}$ into the anticrossing gap: $\Delta_{\bf p}=2\bigl|T^{+-}_{10}\left(\bf p\right)\bigr|\approx \Delta_{0}+\Delta_{\bf p}^{A}$,  where
\begin{eqnarray}
\label{gap-anis}
\Delta_{\bf p}^{A}= \frac{E_{0}f(p)}{\omega_{c}^{e}\Delta_{0}}\Bigl[ C(p_{x}^{2}-p_{y}^{2})+2Dp_{x}p_{y}\Bigl]
\end{eqnarray}
is the anisotropic correction to the gap, and $p$-independent coefficients are given by
%
%\begin{eqnarray}
\begin{align}
\label{CD}
C(\varphi) & =\frac{1}{4}\bigl(\alpha^2+\beta^2\bigr)\sin^2\theta+\alpha\beta\biggl(\cos^4\frac{\theta}{2}+\sin^4\frac{\theta}{2}\biggr)\sin2\varphi,
\nonumber\\
D(\varphi)  & =\alpha\beta\cos\theta\cos 2\varphi.
\end{align}
%\end{eqnarray}
%
Thus, near the resonance, i.e., in a narrow ring in the ${\bf p}$-plane determined by the condition $|\delta_{p}/\Delta_{\bf p}|\lesssim 1$, the MX dispersion is \emph{anisotropic}: $E^{s}_{+}({\bf p})-E^{s}_{-}({\bf p})\sim \Delta_{0}+\Delta_{\bf p}^{A}$. The relative magnitude of the anisotropic energy correction is $\Delta_{\bf p}^{A}/E_{0}\sim (\Delta_{0}/\omega_{c}^{e}) (pl/2)^{2}$; outside of the resonance region, $|\delta_{p}/\Delta_{\bf p}|\gg 1$, anisotropy is negligibly small.

%%%%%%%%%%%%%%%%%%%%%%%%%%%%%%%%%%%%%%%%%
%
\begin{figure}
\centering
\includegraphics[width=0.7\columnwidth]{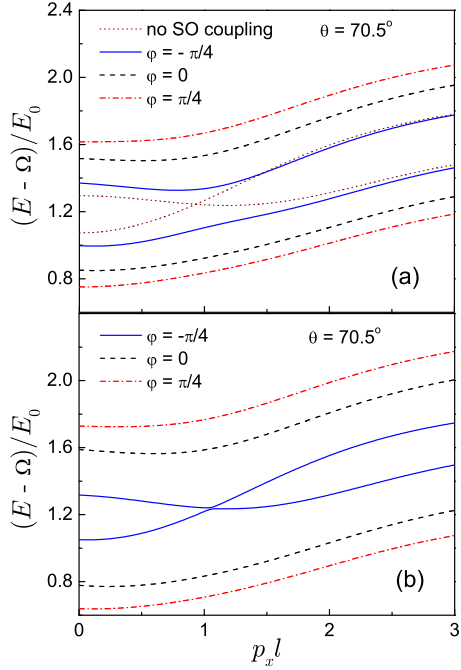}
\caption{\label{fig:dispersion} 
(Color online) Energy dispersion of MX eigenstates $E_{\pm}^{s}({\bf p})$ (upper and lower curves, respectively) at $\beta=157$ meV{\AA} and $\alpha=180$ meV{\AA} (a) and $\alpha=320$ meV{\AA} (b) are plotted for $\theta=70.5^\circ$ and different  $\varphi$. Dotted lines:   dispersions $E_{10}^{++}(p)$ and $E_{00}^{-+}(p)$ in the absence of SO coupling.}
\end{figure}
%
%%%%%%%%%%%%%%%%%%%%%%%%%%%%%%%%%%%%%%%%

In Fig.~\ref{fig:dispersion}, we plot the MX dispersions,  Eq.~(\ref{eigen-energy}), along the $x$-axis of the ${\bf p}$-plane for different in-plane field orientations $\varphi$. The magnitudes of the normal and in-plane field components are taken as $B_{\perp}=4.0$ T and  $B_{\|}=11.3$ T, corresponding to the tilt angle $\theta = 70.5^{\circ}$. For the SO parameters of Fig.~\ref{fig:absorption-1}(a), $\alpha=180$ meV{\AA} and  $\beta=157$ meV{\AA}, the resonance occurs at a finite momentum $pl\sim 1$ [see Fig.~\ref{fig:dispersion}(a)]. At this momentum, the  dispersions $E_{+}^{s}({\bf p})$ and $E_{-}^{s}({\bf p})$ show an anticrossing as $p_{x}$ sweeps through the resonance region, with about factor of 2 gap variation for different $\varphi$. The gap can be strongly reduced by tuning the Rashba coupling $\alpha$, e.g., with the gate voltage.\cite{winkler-book} This is illustrated in Fig.~\ref{fig:dispersion}(b), where the MX dispersions were calculated with larger $\alpha=320$ meV{\AA} that satisfies, at this value of $\theta$, the destructive interference condition, Eq.~(\ref{interference}). It can be seen that the gap practically disappears for $\varphi=-\pi/4$; the effect of coupling to non-resonant LLs (included in the calculation) is undetectable for the chosen parameters.

\begin{figure}[tb]
%\centering
\begin{center}
\includegraphics[width=0.75\columnwidth]{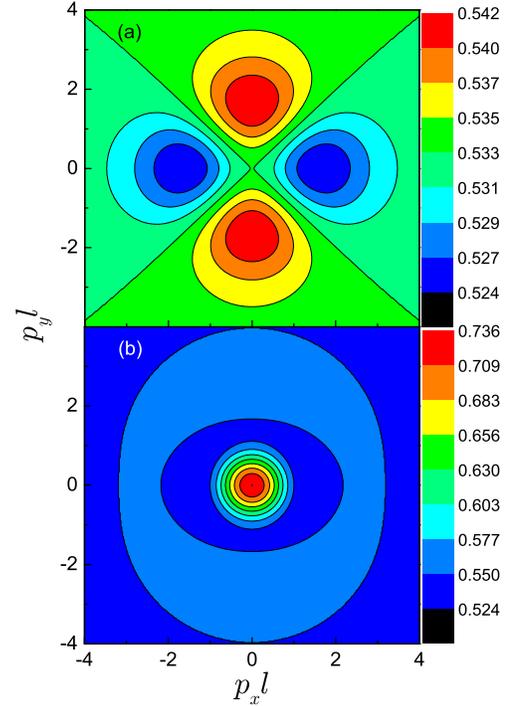}
\caption{\label{fig:anisotropy-1}
(Color online) Contour plots in {\bf p}-plane of $\Delta_{\bf p}/E_{0}$ (a) and $\bigl[E_{+}^{s}({\bf p})-E_{-}^{s}({\bf p})\bigr]/E_{0}$ (b) at $\beta=157$ meV{\AA} and $\alpha=180$ meV{\AA} are shown for  $\theta=69.6^\circ$ and $\varphi=\pi/4$.}
\end{center}
\end{figure}

\begin{figure}[tb]
%\centering
\begin{center}
%\vspace{2mm}
\includegraphics[width=0.75\columnwidth]{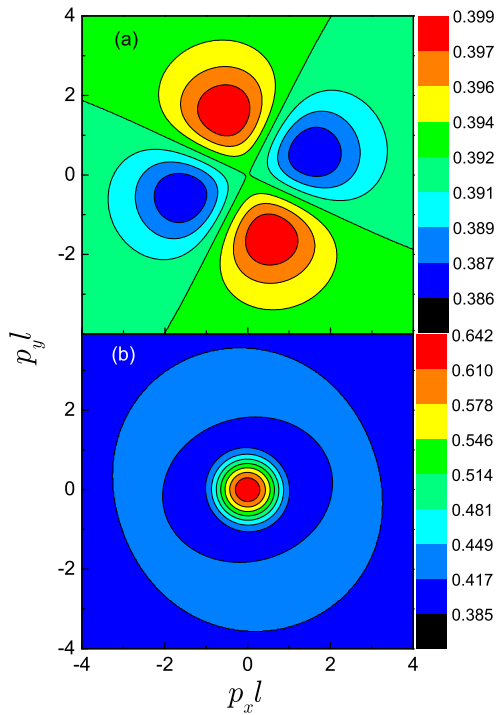}
\caption{\label{fig:anisotropy-2}
(Color online) Same as Fig.~\ref{fig:anisotropy-1} but with $\varphi=0$.}
\end{center}
\end{figure}

In Figs.~\ref{fig:anisotropy-1} and \ref{fig:anisotropy-2}, we show countour plots, in the ${\bf p}$-plane, of the anticrossing gap, $\Delta_{\bf p}$, and MX energy difference, Eq.~(\ref{delta-energy}),  for the in-plane field orientations $\varphi=\pi/4$ and $\varphi=0$. The gap, shown in panels (a), exhibits alternating minima and maxima in a ring-like region $pl= 0.5\div 3.5$ for the chosen SO parameter values [same as in Figs.~\ref{fig:absorption-1}(a) and \ref{fig:dispersion}(a)]; the maximal variation of $\Delta_{\bf p}$ is about 4\% that is comparable to the ratio $(\alpha,\beta)/(l\omega_{c}^{e})$.  For $E_{+}^{s}-E_{-}^{s}$, the anisotropy manifests itself in the elliptical shape of equipotential lines in the ${\bf p}$ plane [panels(b)]. Away from the resonance region, i.e. $\delta_{p}>\Delta_{\bf p}$, the MX spectrum is isotropic. 

%%%%%%%%%%
%
%
%%%%%%%%%%

Note that the anisotropic landscape of MX energy in the ${\bf p}$-plane depends on the in-plane field orientation. For $\varphi=\pm\pi\left/4\right.$, we have $D=0$ in Eq.~(\ref{gap-anis}), so that the extrema of $\Delta_{\bf p}$, as well as the foci of equipotentials, are located on $p_{x}$ and $p_{y}$ axes ($\phi_{\bf p}=0,\pm\pi\left/2\right.$) {\it regardless} of the $\alpha$ and $\beta$ magnitudes [see Fig.~\ref{fig:anisotropy-1} for $\varphi=\pi/4$]. For all other values of $\varphi$, these locations are shifted from $p_{x}$ and $p_{y}$ axes, and the landscape of $E_{\pm}^{s}({\bf p})$ depends on the values of $\alpha$, $\beta$ (see Fig.~\ref{fig:anisotropy-2} for $\varphi=0$). Note finally that the anisotropy of the MX dispersion is more pronounced for  $\varphi=\pi/4$ due to the largest constructive interference between Rashba and Dresselhaus terms for this angle.

%%%%%%%%%%%%%%%%%%%%%%%%%%%%%%%%%%%%%%%
\section{conclusions}
\label{sec:conc}
We have shown that, in a tilted magnetic field, the spin-orbit coupling can significantly change the orbital and spin content of 2D magnetoexcitons. By causing transitions between Landau levels of constituent electrons and holes, SO interaction alters the optical selection rules. This leads to a splitting of the exciton absorption peak when the in-plane field amplitude is tuned to the resonance between bright and dark exciton energies. The splitting magnitude can be varied in a wide range by changing the in-plane field orientation, making possible direct optical measurements of both Rashba and Dresselhaus SO parameters. We also found that the interplay between SO and Coulomb interactions leads to an anisotropy of the exciton energy dispersion that can be, in principle, detected in coupled-QW experiments.\cite{butov-prl01}

Although our consideration was restricted to the lowest LL MXs, the extension to higher LL is straightforward.   In fact, the SO splitting of the exciton absorption peak should be \emph{larger} for higher $n$ due to the larger electronic SO matrix elements, Eq.~(\ref{matrix-all}). It should be noted that, for higher LLs, there are also anticrossings due to other effects of the strong in-plane field such as, e.g., valence band heavy-light hole mixing or Coulomb coupling of LLs from different subbands.\cite{yang_sham-prb87,bauer-prb88,jho-prb05} However, these anticrossings are insensitive to the in-plane field orientation and, therefore, can be easily distinguished from those caused by SO coupling. Finally, this effect is most prominent in narrow-gap semiconductor QWs that are characterized by a strong SO coupling. However, it could be observable in other materials too, e.g., in GaAs where the resonance condition can be achieved with the in-plane field in the range of 60-70 T.

%\acknowledgments
%\section*{Acknowledgments}
This work was supported in part by NSF under Grant No. DMR-0606509 and EPSCOR program, and by DoD under contract No. W912HZ-06-C-0057.

\end{document}